\documentclass[twocolumn,showpacs,preprintnumbers,aps,prb]{revtex4}
\usepackage{graphicx}
\usepackage{dcolumn}
\usepackage{bm}
\usepackage{mathrsfs}
\usepackage{epsfig}
\usepackage{graphicx}
\usepackage{amsmath}
\usepackage{amsfonts}
\usepackage{amssymb}
\usepackage{mathrsfs}
\usepackage{multirow}

\begin{document}

\title{Kink-antikink vortex transfer in periodic-plus-random pinning
potential: \\ Theoretical analysis and numerical experiments} 
\author{W. V. Pogosov$^{1,2}$, H. J. Zhao$^{1}$, V. R. Misko$^{1}$, and F.
M. Peeters$^{1}$}
\affiliation{$^{1}$Departement Fysica, Universiteit Antwerpen, Groenenborgerlaan 171,
B-2020 Antwerpen, Belgium}
\affiliation{$^{2}$Institute for Theoretical and Applied Electrodynamics, Russian Academy
of Sciences, Izhorskaya 13, 125412, Moscow, Russia}
\date{\today }

\begin{abstract}
The influence of random pinning on the vortex dynamics in a periodic square
potential under an external drive is investigated. Using theoretical
approach and numerical experiments, we found several dynamical phases of
vortex motion that are different from the ones for a regular pinning
potential. Vortex transfer is controlled by kinks and antikinks, which
either preexist in the system or appear spontaneously in pairs and then
propagate in groups. When kinks and antikinks collide, they annihilate. 
\end{abstract}

\maketitle


The behavior of an elastic media under the competitive action of a regular
potential and disorder is a common problem in various fields of modern
physics. Examples of such media are vortex lattices in superconductors \cite%
{VVM,Nori,AL,Peeters} or in Bose-Einstein condensates of ultracold atoms 
\cite{Bose-ex}, interacting colloids on periodic substrates\cite%
{Reich,colloids1}, charge and spin density waves in metals \cite{Lub},
polarization density waves in ferroelectrics and many others. Regular
pinning potential can be either of artificial origin, as in nanostructured
superconductors and in Bose-Einstein condensates with optical lattices, or
it can be imposed by the crystal structure of the material. In
superconductors, pinning efficiency determines the value of the critical
current which is of great practical importance. Theoretical description of
such systems is a quite complicated problem, which in one-dimensional case
can be reduced to the well-known Frenkel-Kontorova model 
\cite{Lub,Kivshar}.

Recently, two-dimensional lattice of repelling particles in the presence of
square pinning potential and disorder was investigated theoretically and
numerically \cite{paper2}. In the case of weak disorder, a pinned vortex
lattice is disturbed by specific defects consisting of elastic strings of
depinned vortices. It was found that these strings are able to intersect and
form branched fractal-like clusters, which can percolate through the
system. The aim of the present  
letter is to investigate the vortex dynamics in such a system and look for
correlations between the static defects and the dynamical phases under an
external drive.

\textit{Model.---} We model a three-dimensional (3D) 
superconducting slab by a 2D simulation cell assuming the 
vortex lines are parallel to the cell edges \cite{Nori,AL}, 
in the presence of a regular square array of pinning sites 
with period $a$ and randomly distributed pins of comparable concentration. 
Pinning potential,
produced by a single site, is modeled by a parabolic function, 
with $U_{reg}$
($U_{ran}$) and $\sigma _{reg}$ ($\sigma _{ran}$) being the depth and size
of potential wells of regular (random) origin; $\sigma _{reg}$, $\sigma
_{ran}\ll a$ and $\sigma _{ran}\sim \sigma _{reg}$. We will focus mostly on
the weak-disorder regime, i.e., when the maximum pinning force by one
regular site, $f_{reg}=2U_{reg}/\sigma _{reg}$, is significantly larger than
that for a random site, $f_{ran}=2U_{ran}/\sigma _{ran}$. Vortices are
treated within the London approximation 
(i.e., as 
point-like particles). 
We consider the case of the first matching field, i.e., when the
number of vortices is equal to the number of regular pins. 
To study the motion of vortices, we use molecular-dynamics simulations, and we numerically integrate the overdamped 
equations of motion \cite{Nori,AL}: 
\begin{equation}
\eta \mathrm{\mathbf{v}}_{i}\ =\ \mathrm{\mathbf{f}}_{i}\ =\ \mathrm{\mathbf{%
f}}_{i}^{vv}+\mathrm{\mathbf{f}}_{i}^{vp}+\mathrm{\mathbf{f}}^{d}+\mathrm{%
\mathbf{f}}_{i}^{T}.  \tag{1}
\end{equation}%
Here, $\mathrm{\mathbf{f}}_{i}$ is the total force acting on vortex $i$; $%
\mathrm{\mathbf{f}}_{i}^{vv}$ and $\mathrm{\mathbf{f}}_{i}^{vp}$ are the
forces due to the vortex-vortex and vortex-pin interactions, respectively; $%
\mathrm{\mathbf{f}}^{d}$ is an external driving force (i.e., a Lorentz force
created by an applied current); $\mathrm{\mathbf{f}}_{i}^{T}$ is the thermal
stochastic force. A simulation cell contains $20\times 20$ regular pins,
and we use periodic boundary conditions to simulate an infinite array. To
explain the results of numerical experiments, we also use an analytical
approach: in various situations most of the vortices remain pinned, so that
the vortex transfer occurs through collective defects, which can be
described reasonably well by just few parameters and by using assumptions of
elasticity theory. Below we will present our results for the following set
of parameters: $a=\lambda (T)$, $\sigma _{reg}=0.15a$, $\sigma _{ran}=0.2a$, 
$f_{reg}=0.6f_{0}$, where $f_{0}=\Phi _{0}^{2}/32\pi ^{2}\lambda (T)^{3}$,
numbers of regular and random sites being the same.

\textit{Depinning of stripes (phase I).---} Very weak driving results in no
vortex motion (pinned regime). If the driving force $F_{d}$ reaches some
threshold value $F_{d}^{(I)}$, part of vortices start to move. The vortex
motion is not individual, since vortices travel collectively in a
soliton-like manner, being localized within vortex rows. 
Moving collective structures are just depinned stripe-like defects, which
were predicted in the static configurations\cite{paper2}. Each defect
contains either one extra vortex (kink) or one vacancy (antikink). Kinks
propagate in the direction of an applied force, so that they are solitary
compression waves of vortex row, as seen from Fig. 1 which shows kink
motion. Vacancy-based antikinks flow in the opposite direction; they can be
considered as decompression waves. The number of vortices $D$ in a static
stripe depends on the balance between the vortex-vortex interaction energy
and the energy gained due to the displacement of vortices from the centers
of regular pins 
\begin{equation}
D\approx \left\{ \frac{a}{\lambda (T)}\frac{\left( \Phi _{0}/2\pi \lambda
(T)\right) ^{2}}{2\left( U_{reg}-U_{reg}^{hp}\right) }K_{1}\left[ a/\lambda
(T)\right] \right\} ^{1/2},  \tag{2}
\end{equation}%
where $U_{reg}^{hp}$ is the critical $U_{reg}$, below which the intervortex
interaction dominates over the regular pinning, so that it becomes
energetically favorable for the vortex lattice to switch globally from the
square configuration to the so-called half-pinned phase \cite{paper1,paper2}%
. In this phase, vortices in odd rows remain pinned, while vortices in even
rows are shifted with respect to them over $a/2$, the periodicity of the
lattice being preserved. Eq.~(2) implies that when approaching $U_{reg}^{hp}$%
, the stripe becomes infinitely long. For long stripe, i.e., $D\gg 1$, we
found the criterion for the instability of the pinned stripe, in leading
order in $1/D$. The threshold occurs, when the \textit{total} driving for $D$
vortices in the stripe becomes equal to the resistance force produced by the
first pinned vortex in front of the stripe, which prevents its motion, and
the last pinned vortex behind the stripe, 
\begin{equation}
F_{d}^{(I)}\approx \frac{2f_{reg}}{D}.  \tag{3}
\end{equation}%
Eq.~(3) demonstrates that the stripe behaves nearly as a \textit{rigid} body
containing $D$ particles and this is the explanation why it is depinned at
relatively low driving force, $F_{d}^{(I)}\ll f_{reg}$. Fig.~2 presents
typical numerically calculated average vortex velocity, after steady flow is
achieved, as a function of driving, where one can clearly see distinct
dynamical regions for a weakly disordered regime, while for the stronger
disorder case they are smeared out due to chaotization. Note that in the
particular initial configuration for $f_{rand}=f_{reg}/6$ there was no
stripe in the simulation region, which is reflected by the absence of any
current up to the dynamical phase II (curve 1 in Fig. 2). The results of
Eqs. (2) and (3) are in good agreement with our numerical simulations:
depinning of stripes is predicted to occur at $F_{d}^{(I)}\approx 0.28$ ($%
D\approx 5$, as calculated from Eq. (2)), while in the numerical simulations
this value is around $0.23$ at $f_{rand}=f_{reg}/3$.

\begin{figure}[btp]
\begin{center}
\includegraphics*[width=7.0cm]{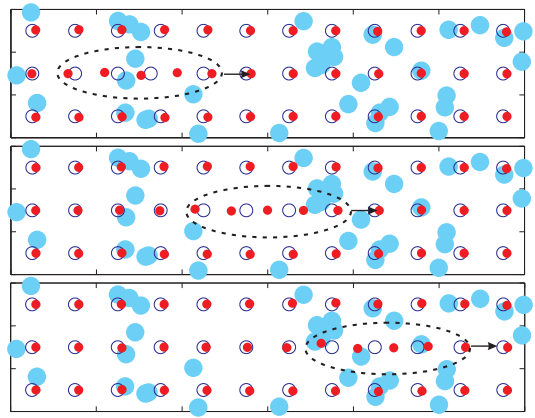}
\end{center}
\par
\vspace{-0.5cm} 
\caption{ (Color online) Motion of the stripe shown by three snapshots.
Irregular blue (light gray) spots represent positions of random pins, red
(dark gray) filled circles show positions of vortices, and regular black
open circles correspond to periodic pins. Dashed lines are guides to the eye
indicating positions of defects, and arrows show the direction of their
motion. }
\label{Fig1}
\vspace{-0.5cm}  
\end{figure}

Kinks and antikinks can collide and \textit{annihilate}. Another remarkable
but quite rare process in the weak disorder regime process is soliton
sticking by bunches of random pins. These two processes lead to a decay of
the total current in time. Fig.~3 shows typical time dependences of the
current, where two examples are addressed, when kinks and antikinks
annihilate (curve 1) and when they persist (curve 1'). It is obvious,
however, that in infinite systems all the kinks and antikinks have to
disappear, since the total number of kinks and antikinks in a single row is
the same. Nonvanishing motion in the weak disorder regime thus appears as an
artifact of a finite-size simulation cell with periodic boundary
conditions.

\begin{figure}[btp]
\begin{center}
\hspace*{-0.5cm} \includegraphics*[width=8.0cm]{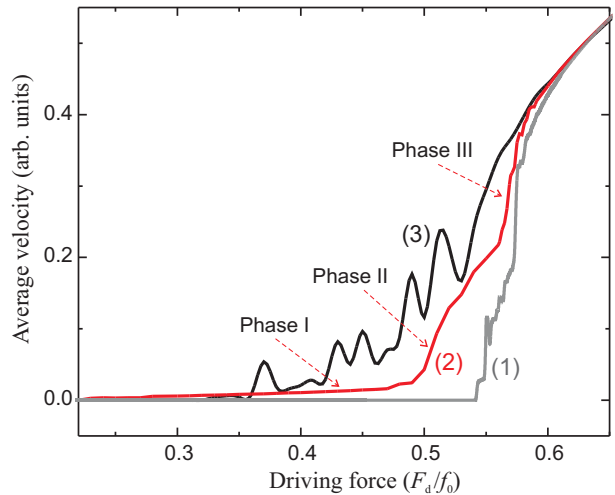} 
\end{center}
\par
\vspace{-0.5cm} 
\caption{ (Color online) The average vortex velocity as a function of
driving for different values of the random pinning force: $%
f_{rand}=f_{reg}/6 $ (curve 1), $f_{reg}/3$ (2), $f_{rand}=f_{reg}/2$ (3). }
\label{Fig2}
\vspace{-0.5cm}  
\end{figure}

\begin{figure}[btp]
\begin{center}
\includegraphics*[width=10.0cm]{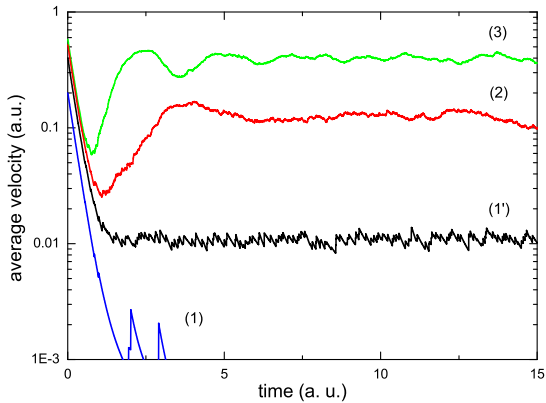}
\end{center}
\par
\vspace{-1.5cm} \vspace{-0.5cm}
\caption{ (Color online) Time dependence of the average vortex velocity for
different dynamical phases at $f_{rand}=0.2$. Here $t_{0}=4\protect\pi ^{2}%
\protect\lambda (T)^{4} \protect\eta / \Phi _{0}^{2}$. Curves 1 and 1'
correspond to phase I at $F_{d}=0.4$ and 0.2, respectively. Curves 2 and 3
show phases II and III for $F_{d}=0.52$ and 0.57, respectively. }
\label{Fig3}
\vspace{-0.5cm}  
\end{figure}

Note that static defects in the weak and intermediate disorder regime
consist not only of stripes. Stripes can be just parts of clusters, some of
which are large and branched \cite{paper2}. Naively one may expect that
these clusters act as easy-channels for vortex transfer. Instead, we found
that low driving partially heals defects. The reason is that clusters
basically consist of two types of segments \cite{paper2}: Segments of the
first kind contain no vacancies or excess vortices, in contrast with
segments of the second type, which are nothing but stripe-like defects. For
low driving, vortices inside segments of the first kind move collectively to
their nearest vacant pins. Such a delicate healing, however, is not possible
for segments of the second type. Therefore, stripes do persist in the sample
and very low driving leads to fragmentation of clusters. A typical driving
force, which heals such defects, can be estimated by considering an infinite
chain of depinned vortices. An effective pinning force for vorticies from
the chain is created by interactions with surrounding rows of pinned
vortices \cite{paper2}. By performing a summation in reciprocal space, we
obtain the following estimate: 
\begin{equation}
F_{d}^{(heal)}\approx \frac{\Phi _{0}^{2}}{2\pi ^{2}\lambda (T)^{2}a}%
e^{-2\pi }.  \tag{4}
\end{equation}%
In the relevant range of parameters, $F_{d}^{(heal)}\ll F_{d}^{(I)}$.

\textit{Generation of kink-antikink pairs (phase II).---} When driving
approaches certain critical value $F_{d}^{(II)}$, kink-antikink pairs are
created spontaneously. New kinks move in the direction of the 
drive, whereas antikinks propagate in the opposite direction. The
corresponding dynamical phase is presented in Fig.~2 by a cusp in curves 1
and 2. Because of the intensive generation of new kinks and antikinks,
average current increases significantly compared to the first dynamical
phase, as seen from Fig.~3. Most of the pair generation events are triggered
by moving stripes in adjacent rows, which create an additional washboard
drive due to vortex-vortex repulsion. Such a process is shown in Fig.~4 by
three snapshots: one of the vortices is depinned, then it creates an area of
vortex row compression in front of itself and the area of decompression
behind, these two areas being transformed into a kink and antikink. It is
easy to realize that the amplitude of the additional periodic drive is equal
to $F_{d}^{(heal)}$ in the limit of long kinks, so the total effective
driving in the row, next to the moving stripe, is $F_{d}^{(eff)}\approx
F_{d}+F_{d}^{(heal)}$. However, a comparison with the numerical results
shows that this value is too low to explain our numerical data: $%
F_{d}^{(heal)}$ is only around 0.03, while the difference between $f_{reg}$
and $F_{d}^{(II)}$ is clearly disorder-dependend: it is about 0.1 for $%
f_{rand}=f_{reg}/3$ and 0.05 for $f_{reg}/6$ (besides, there were no
preexisted kinks within a simulation region in the latter case). The
analysis of vortex motion patterns reveals that kink-antikink pairs are
generated not everywhere. The reason is that there are some weak points,
where vortices are additionally strongly displaced in the direction of drive
by random pins. Concentration of these weak points depends on the external
drive, i.e., on the displacement $r_{0}$ of vortices inside cores, $%
r_{0}\approx \sigma _{reg}F_{d}^{(eff)}/f_{reg}$. Depinning then occurs, if
a pinning force by a random pin is nonzero in this point, which already
restricts the position of the random pin. Another condition is that the
force, produced by the random pin at the edge of the regular pin, together
with \ $F_{d}^{(eff)}$ must dominate the force by the regular site. From
these two conditions, we found a value of the external drive, for which an
appropriate position of a random pin exists 
\begin{equation}
F_{d}^{(II)}\approx \frac{f_{reg}-f_{ran}\left( 1-\sigma _{reg}/\sigma
_{ran}\right) }{1+\left( f_{ran}/f_{reg}\right) \left( \sigma _{reg}/\sigma
_{ran}\right) }-F_{d}^{(heal)}.  \tag{5}
\end{equation}%
Starting from $F_{d}=F_{d}^{(II)}$, a probability for the creation of
kink-antikink pairs is nonzero. We found that above $F_{d}^{(II)}$,
concentration of weak points grows as $\sim (F_{d}-F_{d}^{(II)})^{3/2}$, so
that the kink-antikink pair generation intensifies. Moreover, some pairs
start to nucleate spontaneously in weak points, without the assistance of
moving defects (in particular, for the case of $f_{reg}/6$, when no
preexisted stripes were found within the simulation region). Note that Eq.
(5) was obtained under the assumption that a depinning is due to the action
of a single random pin, without overlap with other ones, such overlaps being
able to produce a larger total pinning force. However, events of this sort
are very rare and they do not smear $F_{d}^{(II)}$ significantly. Also, we
didn't take into account the mutual repulsion of vortices, which acts
against the vortex depinning: according to our estimates, the resistance
force is too small to noticeably change $F_{d}^{(II)}$. Eq. (5) is in a
reasonably good agreement with our numerical results of Fig. 2 for weak
disorder, $f_{rand}=f_{reg}/3$ and $f_{reg}/2$ (in the latter case one has
to drop $F_{d}^{(heal)}$\ from Eq. (5), since there were no preexisted
stripes in our simulation region). In the limit of very weak disorder, Eq.
(5) reduces to $F_{d}^{(II)}\approx f_{reg}-f_{ran}-F_{d}^{(heal)}$, which
clearly shows a competition between regularity and disorder, the latter
factor being enhances by moving kinks through an additional term $%
F_{d}^{(heal)}$. We also notice that, due to a continuous generation of
kink-antikink pairs, vortex transfer within phase II no longer decays in
time, as in phase I. Motion occurs via groups of kinks and antikinks
propagating in opposite directions and creating, from time to time, new
defects, one of which joins the same group, whereas its counterpartner
starts to flow in the opposite direction. Pairs can also be created without
the assistance of previously excited stripes. In steady flow regime,
replication of defects has to be balanced with their annihilation, under
collisions of individual kinks and antikinks or their groups.

\begin{figure}[btp]
\begin{center}
\includegraphics*[width=8.0cm]{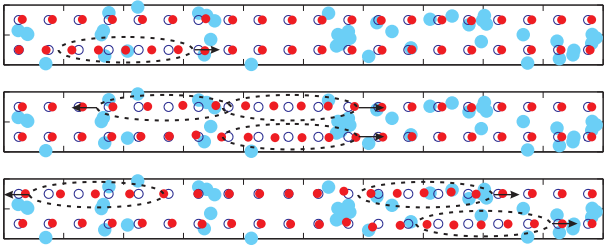}
\end{center}
\par
\vspace{-0.5cm} 
\caption{ (Color online) The same as Fig.~1 but with generation of
kink-antikink pair by moving stripe, shown by snapshots. Initially kink
flows in the bottom row, then it creates a kink-antikink pair in the upper
row. }
\label{Fig4}
\vspace{-0.5cm}  
\end{figure}

\textit{Free proliferation of kink-antikink pairs (phase III).---}
We found third dynamical phase, which shows up as the last distinct region
in curves 1 and 2 in Fig.~2. Careful analysis of numerical data has
indicated that, in this phase, kink-antikink pairs start to proliferate
freely not only in weak points: preexisted kinks immediately generate kinks
and antikinks; in their turn, these defects create new pairs, and so on. 
The area of vortex motion rapidly extends until it covers the
whole simulation region. Mathematically, in this regime, washboard drive
produced by moving kink, is already strong enough to generate kink-antikink
pairs by itself. The corresponding driving is given by 
\begin{equation}
F_{d}^{(III)}\approx f_{reg}-F_{d}^{(heal)}.  \tag{6}
\end{equation}%
In phase III, no \textquotedblleft islands\textquotedblright\ are found,
where vortices can be pinned for a long time. According to Eq. (6), $%
F_{d}^{(III)}$ is 0.57, and this value again is in good agreement with the
numerical results, shown in Fig.~2.

Let us now briefly discuss the dynamics of vortices, when the disorder is no
longer very weak, so that it not only triggers motion of defects, but also
significantly affects it. The general tendency is that disorder smears out
well-separated dynamical phases, as seen from Fig. 2. Stripes can now easily
jump from row to row and bend. However, very surprisingly, soliton-like
origin of the vortex transfer is extremely robust, up to the regime of
strong disorder, $f_{ran}\sim f_{reg}$. In the regime of very strong
disorder, $f_{ran}\gg f_{reg}$, vortex flow is localized in narrow streams,
in which vortices flow one by one. Another mechanism of vortex transfer is
pumping, when vortices are pushed into some traps, until their mutual
repulsion breaks the blockade.

%
%

Note that here we analysed the first-matching-field regime, while a little
imbalance between the numbers of regular pins and vortices could serve as an
additional source of disorder. As was shown in Refs.~\cite{Nori,AL}, this
imbalance results in different dynamical regimes including vortex flow in
``incommensurate rows'' and negative-differential-resistivity (NDR) parts of
the VI-curve of N- \cite{Nori,AL} and S-type \cite{AL}. Very recently, the
first experimental observation of the N-type NDR phase has been reported 
\cite{VVM1}. (See also a related experiment \cite{Koelle} on a triangular
array of pins, where channeling of vortices can be suppressed by the random
removal of pinning sites \cite{Reich-1}.)

\textit{Conclusions.---} We studied the competitive effect of periodic
square and weak random pinning potentials on the dynamics of vortices in two
dimensions. We found new dynamical phases, which are established through a
cascade of transitions. There are three phases, in which vortices move in a
soliton-like collective structure travelling within individual vortex rows.
These are kinks, each containing an excess vortex and moving in the
direction of an external drive, and antikinks, flowing in the opposite
direction and containing a vacancy. When colliding, kinks and antikinks
annihilate. Initial motion is triggered by depinning of preexisted static
kinks and antikinks. In the second regime, moving defects generate secondary
kink-antikink pairs in adjacent rows at certain weak points, which are more
corrupted by disorder. In the third regime, these pairs are freely generated
by moving kinks and antikinks, i.e., without help of disorder.

Although we have concentrated on vortices in superconductors, it is clear
that similar dynamical regimes will be realized in other two-dimensional
systems with periodic lattice potentials, containing repelling particles.
Moreover, disorder-induced kink-antikink generation under an external drive
can appear as a rather universal phenomenon, which exists for systems and
lattice potentials of various dimensionalities.

This work was supported by the ``Odysseus'' Program of Flemish government,
FWO-Vl, and IAP. W.V.P. acknowledges support from RFBR (contract
No.~06-02-16691). V.R.M. acknowledges support from the EU MC Programme,
Contract No.~MIF1-CT-2006-040816.

\end{document}